\documentclass[aps,prb,manuscript]{revtex4}
\usepackage{graphicx}
\usepackage{color}
\usepackage{bm}
\usepackage{float}

\begin{document}
\draft

\title{Influence of electric and magnetic fields and $\sigma$-edge bands on the electronic and optical spectra of graphene nanoribbons}

\author{Thi-Nga Do$^{1}$, Po-Hsin Shih$^{2}\footnote{Corresponding author: {\em E-mail}: phshih@phys.ncku.edu.tw}$, Godfrey Gumbs$^{3}$, Danhong Huang$^{4}$}
\affiliation{$^{1}$ Center for High Technology Materials, University of New Mexico, Albuquerque, New Mexico 87131, USA \\
$^{2}$Department of Physics, National Cheng Kung University, Taiwan 701\\
$^{3}$Department of Physics and Astronomy, Hunter College of the City University of New York, 695 Park Avenue, New York, New York 10065, USA \\
$^{4}$US Air Force Research Laboratory, Space Vehicles Directorate (AFRL/RVSU), Kirtland Air Force Base, New Mexico 87117, USA
}

\date{\today}

\begin{abstract}
The unusual electronic and optical properties of armchair and zigzag graphene nanoribbons (GNRs) subject to in-plane transverse electric and perpendicular magnetic fields have been systematically investigated. Our calculations were carried out within the generalized multi-orbital tight-binding model based on a Hamiltonian which takes into account hopping integrals among the (s, $p_x$, $p_y$, $p_z$) atomic orbitals as well as the external electric and magnetic fields. The electronic structure consists of $\pi$ bands arising from the $p_z$ orbital and $\sigma$ bands originating from  the (s, $p_x$, $p_y$) orbitals. The energy bands and optical spectra are diversified by both the nature of the edge of the nanoribbon and strength of the external fields. Armchair GNRs display a width-dependent energy gap in addition to low-energy $\sigma$ bands while the zigzag system has the unfilled flat band with $\pi$ edge states at zero energy and partially filled wide-range $\sigma$ bands. An applied in-plane electric field leads to the splitting of energy bands and shifted Fermi level, thereby enriching the inter-band and intra-band optical conductivities. The interplay between an external magnetic field and the edge geometry gives rise to extraordinary quantized Landau levels and special optical spectra.
\end{abstract}
\pacs{PACS:}
\maketitle

\section{Introduction}
\label{sec1}

Graphene nanoribbons, which are narrow strips of graphene, have been receiving a considerable amount of attention due to their acquired fundamental physical properties as well as their wide range of potential applications. Unlike gapless graphene, GNRs open up band gaps as a result of quantum confinement and edge effects\,\cite{gapopen1, gap2, gap3, aczz}. The quasi-one-dimensional nature of GNRs plays a critical role in the exceptional characteristics, making the materials appropriate for use in nanoelectronics, optoelectronics, spintronics, photodetectors, quantum devices and others\,\cite{gap3, electronic1, electronic2, optoelectronic, spintronic, detector1, detector2, device}. So far, GNRs have been successfully synthesized by various methods, including lithography\,\cite{lithography1, lithography2, lithography3, gapopen1}, bottom-up\,\cite{bottomup1, bottomup2, bottomup3}, as well as unzipping carbon nanotubes\,\cite{unzipping1, unzipping2, unzipping3}.
It is worth mentioning that the current experiment techniques to synthesize GNRs, such as
top-down and unzipping, might miss  control over the edge passivation\,\cite{bottomup2, passivationexp1, passivationexp3}.
GNRs passivated with other atoms, such as hydrogen, oxygen, and metals on the edge, will alter their fundamental properties,
as already demonstrated by various research groups\,\cite{passivation, 4orbital}.
It is desired to investigate the electronic and optical properties of GNRs without removing the $\sigma$-edge bands by the passivation.
In our study of GNRs, we consider the pristine zigzag and armchair GNRs, and compare our results with those reported previously for both passivated and nonpassivated GNRs.
Our aim is threefold: (i) to comprehend the effects of an in-plane transverse electric field on the electronic and optical properties, (ii) to explore the rich and unique orbital quantization phenomena, one of the mainstream topics in physical science, and (iii) to thoroughly investigate the diversified magneto-optical excitations accompanied by specific selection rules.
\medskip

Up to now, a large number of theoretical and experimental studies on the properties of GNRs have been performed. Especially, the electronic and optical characteristics of GNRs have been demonstrated to be remarkably diversified by the ribbon width and edge types\,\cite{gapopen1, gap2, gap3, aczz, lithography3, optical}. Theoretical calculations have suggested that the electronic structure of GNRs displays the width-dependent energy gaps for armchair terminations and partially flat bands with edge states for zigzag structures\,\cite{gapopen1, gap2, gap3, aczz, optical}. Magnetic quantization is predicted to be significantly suppressed by lateral confinement; Landau levels (LLs) compete with quantum confinement and are only observed for sufficiently wide GNRs\,\cite{moptical, field1}. The optical-absorption spectra are sensitively affected by both the electric and magnetic fields, in terms of spectral structure, intensity and frequency\,\cite{optical,moptical,field1}. From an experimental point of view, opening of energy gaps, the edge states and their dependence on the lateral confinement have been verified through measurements of temperature dependent conductance in the nonlinear response regime\,\cite{gapopen1}, room-temperature on-off current switching\,\cite{current}, by scanning tunneling microscopy as well as by scanning tunneling spectroscopy\,\cite{sts}. Optical measurements have been conducted for GNRs\,\cite{bottomup1, opexp1, opexp2}, in which the optical gap and the geometry-dependent shifts of the absorption peaks are examined.
\medskip

However, previous calculations on the electronic and optical properties of GNRs under electric and magnetic fields were limited to the tight-binding model (TBM) with only one $p_z$ orbital per atom\,\cite{optical, moptical, field1}. Apparently, such a simple model is not able  to capture the full extent of the dynamics of the band structure and optical-absorption spectra due to the lack of critically dispersive $\sigma$ bands connected with the shift in the  Fermi level.
So far, the four-orbital energy bands, including the $\pi$ bands made of $p_z$ and $\sigma$ bands made of (s, $p_x$, $p_y$), have been reported in the literature.\cite{4orbital, firstp}  In addition to the $\pi$ bands, the $\sigma$ bands located near $E = 0$ are also important in our understanding of the low-energy physics of GNRs. As far as we know, there is still  insufficient study of the magnetic quantization and magneto-optical properties of four-orbital GNRs. Consequently, this topic deserves a careful investigation.
\medskip

Motivated by recent theoretical and experimental progress on these materials, we have explored the role played by electric and magnetic fields on the electronic and optical properties of four-orbital armchair and zigzag GNRs. The interplay between the external electric and magnetic fields and the edge geometry yields distinctive band structures and LL features, giving rise to peculiar inter-band and intra-band optical conductivities. We will show that, the external electric and magnetic fields can separate the doubly degenerate $\sigma$-edge bands differently, in addition to the edge passivation as reported previously.
An $E$ field evidently modifies the absorption spectra through the shift of Fermi level. On the other hand, a $B$ field can only change the low-frequency spectral structures related to the weakly quantized LLs. The interesting  field-induced  energy dispersion and the optical transitions associated with them will be clearly discussed. The comparison between our results and the previous theoretical and experimental reports will be carried out.

\medskip

The rest of this paper is organized as follows. In Sec. \ref{sec2}, we describe our generalized TBM which we used for calculating the energy band structure for GNRs with armchair and zigzag edges. Section \ref{sec3} is devoted to numerical calculations and discussion of the electronic and optical properties of GNRs with armchair and zigzag edges  in the absence and presence of an in-plane electric field as well as a perpendicular magnetic field. We summarize our results in Sec. \ref{sec4}.
\medskip

\section{Method}
\label{sec2}

We have developed the multi-orbital nearest-neighbor TBM to investigate the electronic and optical properties of GNRs with armchair and zigzag edges in an electric and a magnetic fields. GNRs are composed of two equivalent sublattices, referred to as A and B, as shown in Figs. 1(a) and 1(b) for armchair and zigzag edges, respectively. We choose $x$ and $y$ directions for transverse and longitudinal directions with respect to a nanoribbon, respectively. The primitive unit cells, marked by the red rectangles, consist of 2N carbon atoms where $N$ is the number of armchair or zigzag lines.
The first Brillouin zone is determined by the requirement that $k_y$ is within [$-\pi/L_y$, $\pi/L_y$], where $L_y$ is the length  of the periodic primitive unit cell. The nanoribbon widths of the armchair and zigzag GNRs are defined as $W_{ac} = L_y(N-1)/2$ ($L_y = 3b$ with $b$ = 1.42 \AA \space being the C-C bond length) and $W_{zz} = L_y(N/2-1/3)$ ($L_y = \sqrt{3}b$), respectively.
\medskip

The Hamiltonian, including the $sp_2$ orbital bonding and an external electric field, is given by\,\cite{tbm1,tbm2}

\begin{equation}
\hat{H} = \sum_{\langle i \rangle, o} (\epsilon_o + {\cal V}_{sc}(x_j)) \hat{C}_{io}^{\dag} \hat{C}_{io}
+ \sum_{\langle i,j \rangle, o,o^{'}} t_{oo^{'}}^{R_{ij}} \hat{C}_{io}^{\dag} \hat{C}_{jo^{'}}.
\end{equation}
In this Hamiltonian, the $\hat{C}_{io}^{\dag}$ ($\hat{C}_{jo^{'}}$) operator could create (annihilate) an electronic state with orbital $o$ ($o^{\prime}$) at lattice site $i$ ($j$). Also, $\epsilon_o$ is the orbital-dependent on-site energy, $t_{oo^{\prime}}^{R_{ij}}$ is the nearest-neighbor hopping integral which depends on the two atomic orbitals of ($o$, $o^{\prime}$) and the translation vector $R_{ij}$ between two atoms.
It is crucial to mention that applying a voltage drop across the nanoribbon does not ensure a spatially-homogeneous electric field in the system. Instead, a field domain will be induced self-consistently. By neglecting insignificant screening contribution from edge-state electrons, we approximate the screening by a static dielectric function $\epsilon_{s}(q_x,q_y)$ of quasi-one-dimensional graphene ribbon. The statically-screened potential can be written as [38]

\begin{equation}
	{\cal V}_{sc}(x_j)\equiv{\cal V}_{sc}(x_j,y=0)={\cal A}_0\,{\rm Re}\left\{\,\int\limits_{-\infty}^{\infty} dq_x\,\texttt{e}^{iq_xx_j}\,U_{ext}(q_x)\,\int\limits_{-\infty}^{\infty} \,\frac{dq_y}{\epsilon_{s}(q_x,q_y)}\right\}\ ,
	\label{e1}	
\end{equation}
where the dimensionless constant ${\cal A}_0$ is fixed by the constraint
${\cal V}_{sc}(W)-{\cal V}_{sc}(0)=-EW$ in which E represents the applied uniform electric field across the width of a nanoribbon, and $W$ is the nanoribbon width.
Here, different choices of $y$ value only gives rise to a phase factor.
In addition, $U_{ext}(q_x)$ is the Fourier-transformed external potential $v_{ext}(x_j)=-eEx_j$, given by

\begin{eqnarray}
	\nonumber
	U_{ext}(q_x)&=&\int\limits_{0}^{W} dx'\,\texttt{e}^{-iq_xx'}\,v_{ext}(x')
	=-\frac{ieE}{q_x^2}\left[ iq_xW \texttt{e}^{-iq_xW} +  \texttt{e}^{iq_xW} -1 \right]\ .
\end{eqnarray}

The dynamical dielectric function $\epsilon_{2D}(q_{xy},\omega)$ is calculated from the random-phase approximation as [37]

\begin{eqnarray}
	\nonumber
	\epsilon_{2D}(q_{xy},\omega) &=& \epsilon_b - \nu_{2D}(q_{xy})\,\left|{\cal F}(q_x)\right|^2\,\sum\limits_{c,v}\,\int\limits_{1stBZ} 2\frac{dk_y}{2\pi} \left| \Big\langle  k_y + q_y;c \left| \texttt{e}^{iq_yy}\right| k_y; v\Big\rangle \right|^2\\
	&\times&
	\frac{f(E_{c}(k_y +q_y))- f(E_{v}(k_y))} {E_{c}(k_y +q_y)- E_{v}(k_y) -(\omega + i\delta)}\ ,
	\label{e2}	
\end{eqnarray}
where $\epsilon_b$ = 1 is the background dielectric constant, $\nu_{2D}(q_{xy})=e^2/(2\epsilon_0q_{xy}W)$ is the bare potential,
$q_{xy}=\sqrt{q_x^2+q_y^2}$, and ${\cal F}(q_x)$ represents the dimensionless form factor which can be computed by

\begin{eqnarray}
	\nonumber
{\cal F}(q_x)=\int\limits_{0}^{W} dx\,\left|\psi(x)\right|^2\,\texttt{e}^{iq_xx}
\end{eqnarray}
with $\psi(x)$ as the transverse envelope function for a nanoribbon obtained by integrating over $y$. Moreover, we have $\epsilon_{s}(q_x,q_y)=\epsilon_{2D}(q_{xy},\omega=0)$.

The parameters used in our calculations are optimized numerically, following Ref.\ [\onlinecite{parameter}], so as to reproduce the energy bands calculated previously by the first-principles method and TBM\,\cite{4orbital, firstp}. The application of an external perpendicular magnetic field is included in the calculations by adding an extra position-related Peierls phase in the nearest-neighbor hopping integral\,\cite{moptical, absorption}.
\medskip

When GNRs are irradiated by an electromagnetic field, there exist vertical optical excitations from occupied to unoccupied states. The finite intensity of such excitations could be determined from the absorption function\,\cite{absorption},

\[
A(\omega) \propto \frac{1}{(2\pi)^2}
\sum\limits_{c,v}\sum\limits_{m,m'}\,\int\limits_{1stBZ} d^2\mbox{\boldmath$k$}
 \left| \Big\langle \Psi^{c} (\mbox{\boldmath$k$},m')
 \left| \frac{   \hat{\mathbf{E}}\cdot \mathbf{P}   } {m_e}
 \right| \Psi^{v}(\mbox{\boldmath$k$},m)    \Big\rangle \right|^2
 \]
\begin{eqnarray}
\times
Im \left[      \frac{f_0(E^c (\mbox{\boldmath$k$},m')) - f_0(E^v (\mbox{\boldmath$k$},m))}
{E^c (\mbox{\boldmath$k$},m')-E^v (\mbox{\boldmath$k$},m)-\omega - i\Gamma}           \right],
\end{eqnarray}
where $f_0(x)=\Theta(E_F-x)$ with Fermi energy $E_F$, $\Theta(x)$ is the unit-step function, $\Big\langle \Psi^{c} (\mbox{\boldmath$k$},m')
\Big| \frac{   \hat{\mathbf{E}}\cdot \mathbf{P}   } {m_e}
\Big| \Psi^{v}(\mbox{\boldmath$k$},m)    \Big\rangle$ is the velocity matrix element, and $Im \Big[\frac{f_0(E^c (\mathbf{k},m')) - f_0(E^v (\mathbf{k},m))} {E^c (\mathbf{k},m')-E^v (\mathbf{k},m)-\omega - i\Gamma}   \Big] $ is the joint density of states. $\Gamma$ is the lifetime broadening factor which is chosen to be sufficiently small for free-standing systems ($\Gamma$ = 1 meV). Previous work shows that this method could yield highly accurate optical absorption spectra which are consistent with experimental results\,\cite{absorption}.
\medskip

\section{Results and Discussion}
\label{sec3}

\begin{figure}[h]
\centering
{\includegraphics[width=0.8\linewidth]{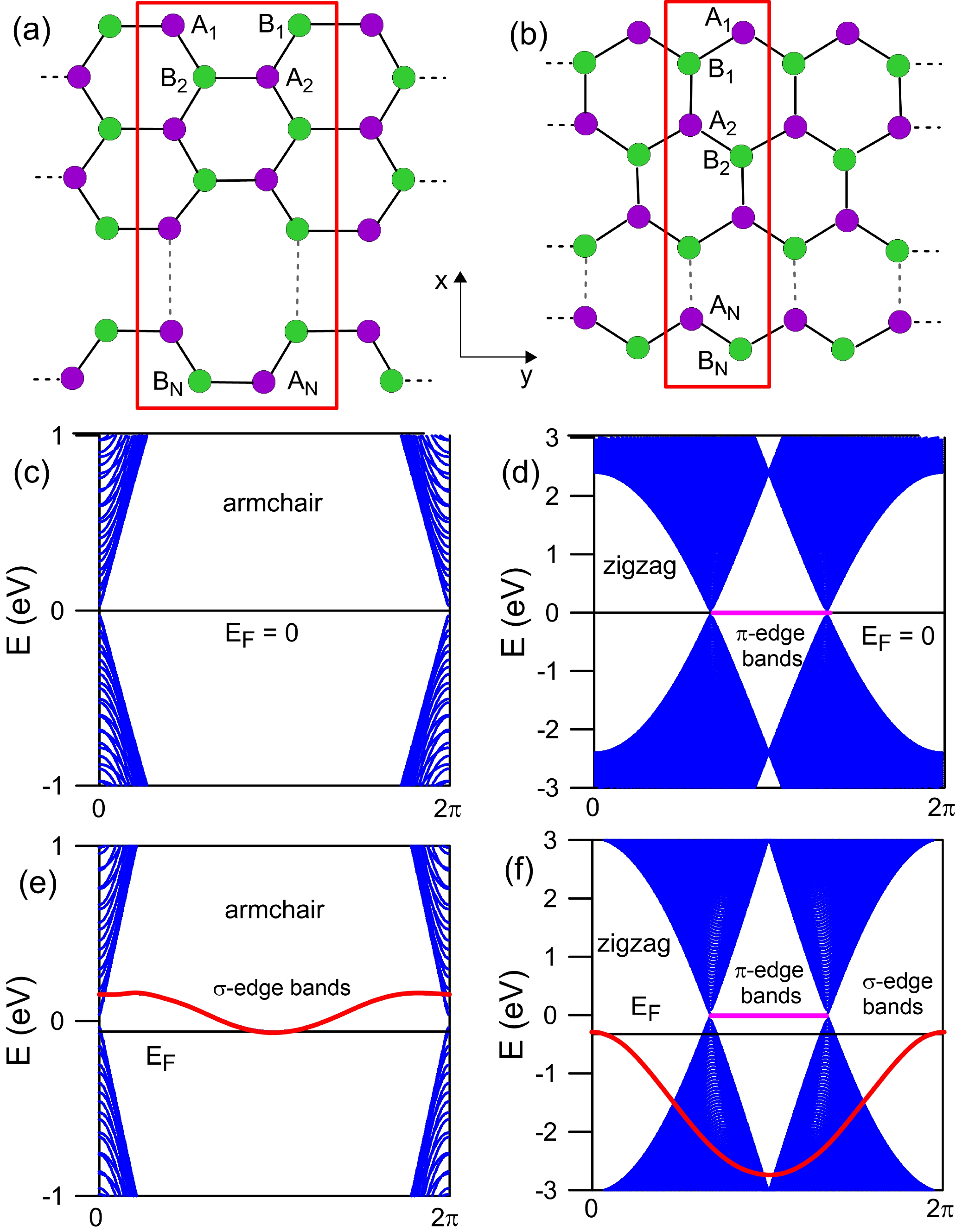}}
\caption{(color online) Lattice structures of GNRs with (a) armchair and (b) zigzag edges for $N = 150$. The calculated energy bands for both $p_z$ and (s, $p_x$, $p_y$, $p_z$)-orbitals TBMs are presented in (c), (d) and (e), (f), respectively.}
\label{Fig1}
\end{figure}

The band structures of armchair and zigzag GNRs calculated with the use of both $p_z$ and multi-orbital TBMs are presented in Figs. \ref{Fig1}(c) through \ref{Fig1}(f) for comparison. The $p_z$ orbital TBM gives only the $\pi$ bands, as demonstrated in Figs. \ref{Fig1}(c) and \ref{Fig1}(d). The low-lying energy dispersion forms the parabolic shapes and it varies with the ribbon edges. The conduction and valence bands are symmetric about the Fermi level $E_F = 0$.  As for the zigzag GNRs, a flat band appears at the Fermi level within the  range 2$\pi$/3 $\leq$ $k_y$ $\leq$ $\pi$. On the other hand, the armchair structure exhibits a width-dependent band gap between the edge bands at $k_y = 0$ which tends to zero as $N$ is increased. It has been predicted by the $\pi$-bands TBM and also the first-principle calculations that an armchair GNR is semiconducting except for $N = 3p +2$ with $p$ as a positive integer, where it becomes metallic.\,\cite{gapopen1, gap2, gap3, aczz}  The metallic behavior of an armchair GNR has also been examined by experimental measurements.\,\cite{acgapexp}  Here, the $\pi$ band structures in our model study of GNRs with $N = 150$ are consistent with those obtained in the previous works.\cite{gapopen1,gap2, gap3, aczz, optical, moptical}
\medskip

Since the lattice symmetry is broken at the ribbon edges, the multi-orbital Hamiltonian is necessary instead of just $p_z$ one. As a matter of fact, the $\sigma$-edge bands come into existence when the four (s, $p_x$, $p_y$, $p_z$) orbitals are included in the calculations.\,\cite{4orbital}  Interestingly, the $\sigma$ energy bands are well separated from the $\pi$ bands. They are mainly made of (s, $p_x$, $p_y$) orbitals, unlike the $\pi$-edge bands which consist of only $p_z$ orbital. Figs. \ref{Fig1}(e) and \ref{Fig1}(f) present the low-lying energy bands of armchair and zigzag GNRs, respectively. The weak energy dispersion of $\sigma$ bands occurs near zero energy for armchair GNRs while its strong dispersion enters into much deeper energy for the zigzag system. The $\sigma$ bands are doubly degenerate, corresponding to two identical ribbon edges. The relative position of the edge $\sigma$ and $\pi$ bands determines the Fermi level, whereby $E_F$ is located above or below the $\pi$ bands for the armchair and zigzag GNRs, respectively. Our numerical calculations show that $E_F  = -0.0615$ eV for armchair and $E_F = -0.3289$ eV for zigzag terminations with N = 150. Interestingly, the electronic characteristic of armchair GNR is gradually changed from semiconducting to metallic when the ribbon width increases. We have confirmed that the armchair GNRs with N $\leq$ 30 presents the semiconducting behavior, in consistent with the previous reports\,\cite{gapopen1, gap2, gap3, aczz}. On the other hand, the N = 150 armchair GNR becomes metallic as the Fermi level crosses the energy bands. This finding might be an important reference for the future experimental verification.
\medskip

\begin{figure}[h]
\centering
{\includegraphics[width=0.5\linewidth]{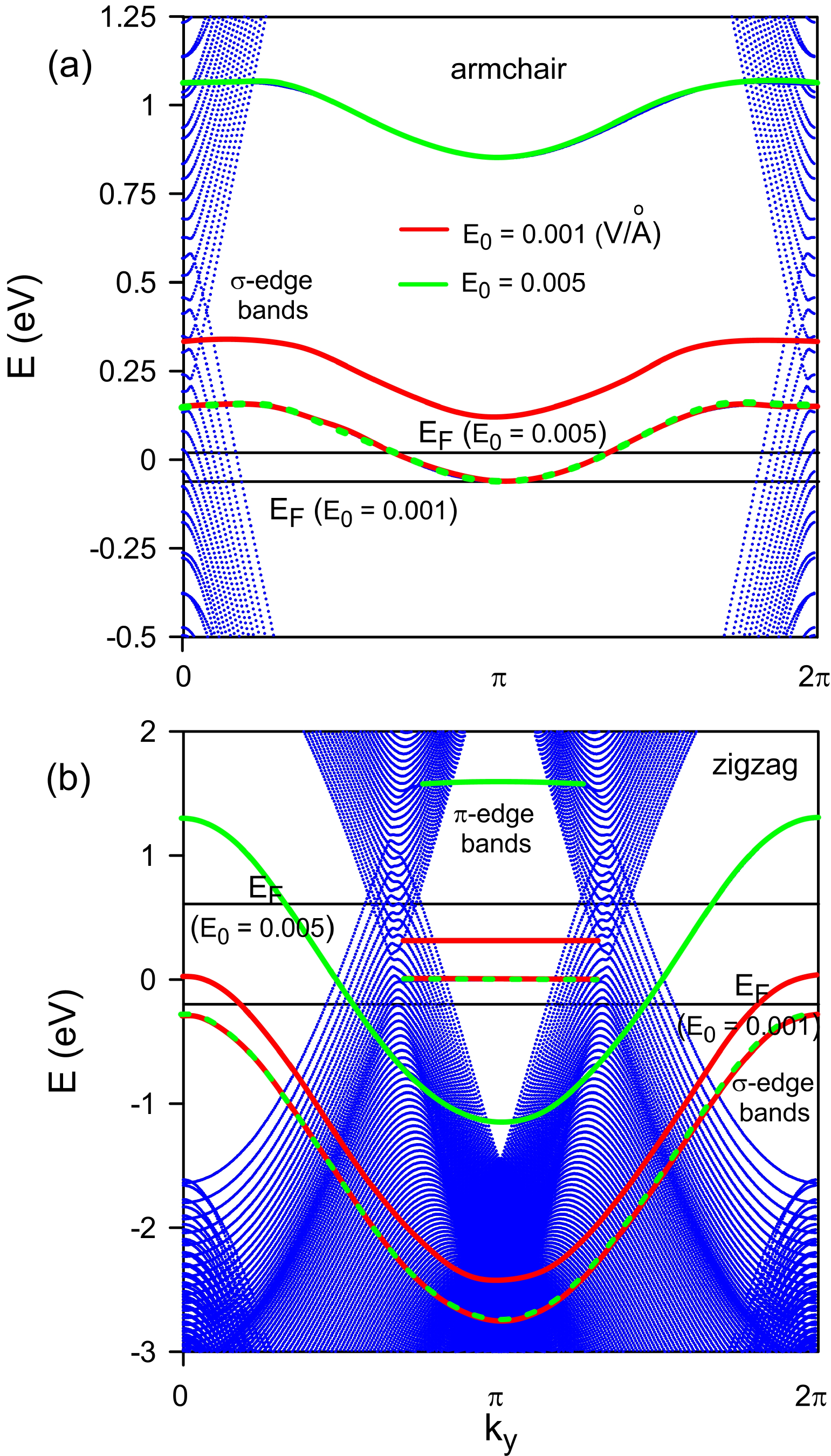}}
\caption{(color online) Calculated energy bands of (a) armchair and (b) zigzag GNRs using the four-orbitals TBM in the presence of an external in-plane transverse electric field $E_0$ which generates a band splitting between two degenerate $\sigma$ bands. The dotted red and green lines represent the edge bands which remain unchanged under the application of $E_0$.}
\label{Fig2}
\end{figure}

When an in-plane transverse electric field is applied across the ribbon edges, the screened potential causes the significant modification of the electronic structures, including the energy dispersion and the band splitting. Figures \ref{Fig2}(a) and \ref{Fig2}(b) show the band structures with a finite $E$ field of the armchair and zigzag GNRs, respectively. The field conspicuously narrows the separation between the conduction and valence $\pi$ bands for both systems. The influence due to the $E$ field is more visible for the $\pi$- and $\sigma$-edge bands. One of the two degenerate edge bands remains unchanged whereas the other is shifted upward which as a result of the band splitting. Such an energy splitting is uniform along $k_y$ and it becomes wider when the field is increased. The effect due to the electric field on the band structure gives rise to a shift in the Fermi level. We observe that $E_F$ is shifted more upward for larger field, which is consistent with the electric field-dependent energy dispersion. We also note that the presence of the $\sigma$-edge bands strengthens the effect of the $E$ field on variation of the Fermi level.
It is noticed that the feature of band structures under an electric field is quite different from the passivated GNRs in which the hybridization between the $\sigma$-edge bands near the zero energy and the passivated atomic orbitals gives rise to the separation between the original and newly introduced edge bands.\,\cite{4orbital}
The $E$-field-induced rich electronic structures significantly alter the vertical optical transitions from the occupied to the unoccupied states which we will discuss next.
\medskip

\begin{figure}[h]
\centering
{\includegraphics[width=0.7\linewidth]{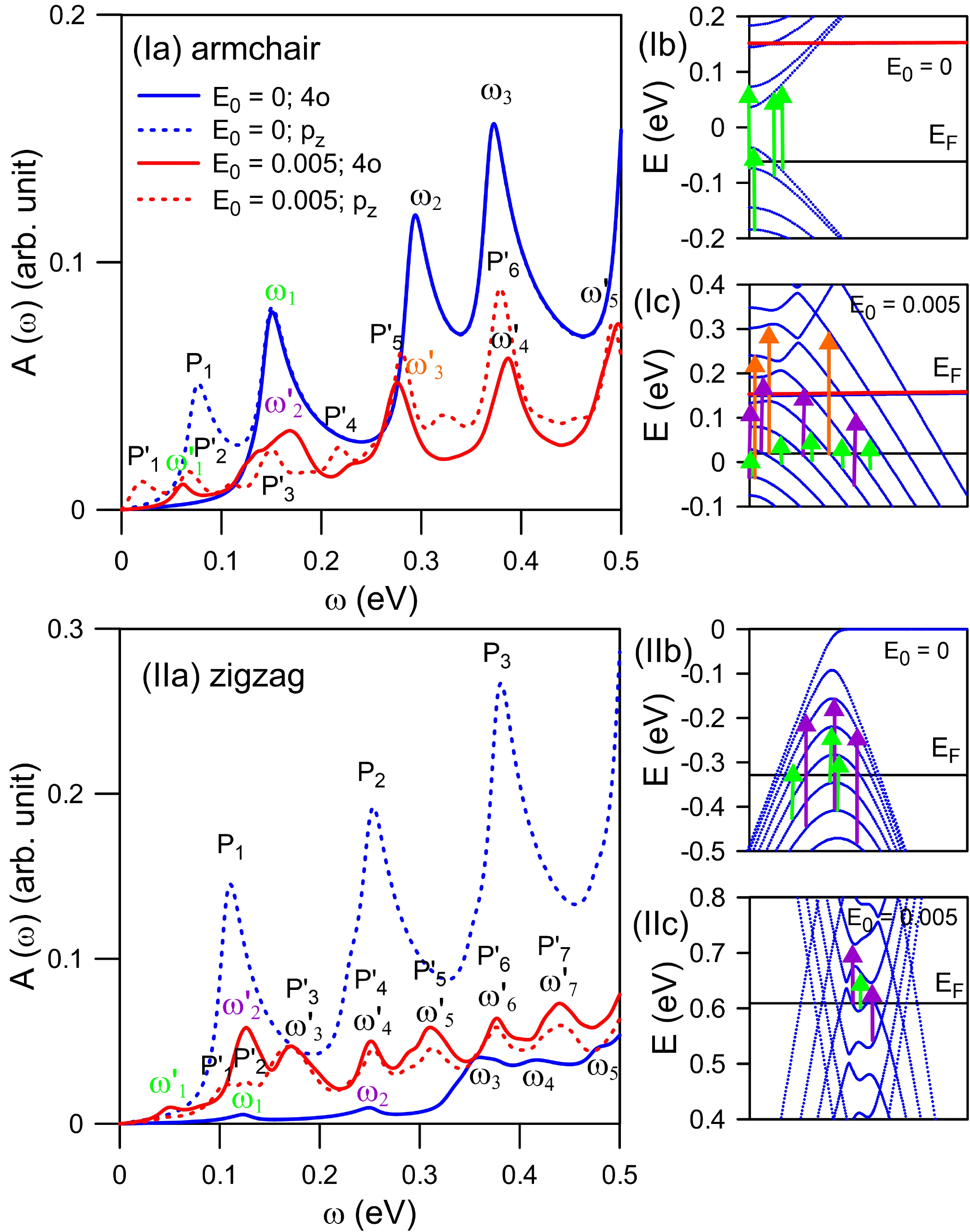}}
\caption{(color online) Calculated absorption spectra of (Ia) armchair GNR with and without an $E$ field (in V/$\AA$). Results for $p_z$ and (s, $p_x$, $p_y$, $p_z$)-orbital TBMs are represented by the dashed and solid curves, respectively. The corresponding vertical transition channels are shown in (Ib) and (Ic). Similar plots for zigzag edge GNR are presented in (IIa) through (IIc).}
\label{Fig3}
\end{figure}

The optical absorption function exhibits peak and shoulder-like structures, corresponding to vertical transitions between band-edge states or the multi-excitation channels. The characteristics of the  spectral structure strongly depend on the lateral confinement as well as the frequency range, referring to the blue solid curves in Figs. \ref{Fig3}(Ia) and \ref{Fig3}(IIa) for the armchair and zigzag GNRs at zero field, respectively. For the armchair edge, the pronounced peaks correspond to vertical transitions from the occupied valence to unoccupied conduction band states, as illustrated in Fig. \ref{Fig3}(Ib). The vertical green arrows indicate the excitations forming the threshold peak $\omega_1$. The four-orbital TBM yields similar spectral structures compared with the single $p_z$-orbital TBM (the dashed blue curve in Fig. \ref{Fig3}(Ia)) except for the disappearance of the $P_1$ excitation. In fact, the $P_1$ peak corresponds to the vertical transition from the highest valence state to the lowest conduction one, its frequency measures the finite band gap of the armchair system. The lack of such a transition by including the four orbitals is because of the emergence of the $\sigma$-edge bands which repositions the Fermi level. The zigzag-edge GNR displays remarkable differences between the absorption spectra correlated with the single-orbital and multi-orbital model calculations. By including the (s, $p_x$, $p_y$, $p_z$) orbitals, the Fermi level is significantly lowered toward the valence bands. Therefore, the weakened absorption peaks in the low frequency range are associated with the multi-channel vertical transitions among the valence states, as demonstrated in Fig. \ref{Fig3}(IIb). All of these peaks are not well separated since the closeness of electronic states in the vicinity of $E_F$ gives rise to plenty of excitation channels with only slight difference in frequency. This leads to the emergence of shoulder-like spectral structures, such as $\omega_3$ in Fig. \ref{Fig3}(IIa). It is worth noting that, the absorption spectra of both the single $p_z$-orbital and the four-orbital Hamiltonian matrices using the Fermi energy obtained by the four-orbital TBM are equivalent. This is consistent with the optical selection rule in which the vertical transition between the two $\sigma$-edge bands are forbidden.
\medskip

The influence of finite electric field on the optical-absorption spectra is mainly attributed to the shift of the Fermi level and the distortion of band edge states. The solid red curves in Figs. \ref{Fig3}(Ia) and \ref{Fig3}(IIa) illustrate the spectral structures of (s, $p_x$, $p_y$, $p_z$)-orbital armchair and zigzag GNRs for $E_0 = 0.005$ V/\AA.  An electric field brings out  remarkable changes in the spectra, including  the alteration of the frequency and peak intensity, as well as the enhancement of shoulder-like structures. The threshold structures $\omega^{\prime}_1$ of both armchair and zigzag GNRs arise at lower frequency compared with those for the zero-field spectra. They are associated with the vertical transitions of the dense electronic states around the new Fermi levels, as demonstrated by the vertical green arrows in Figs. \ref{Fig3}(Ic) and \ref{Fig3}(IIc). For armchair GNRs, the low frequency spectral structures are formed by excitations within the valence bands, differing from  the above mentioned zero-field spectrum. Furthermore, the band edge states of the parabolic bands are deformed or anti-cross due to band coupling, as shown in Fig. \ref{Fig3}(Ic). These are responsible for the lowering of spectral intensity and the emergence of shoulder-like structures interspersed among the peaks. For zigzag GNR, the low frequency absorption spectrum is correlated with the transitions among the valence bands regardless of whether an electric field is applied or not. However, the Fermi level under a field enters into deeper valence bands where the density of states becomes much higher. As a result, the spectral intensity is increased by an $E$ field, as shown in Fig. \ref{Fig3}(IIa), in contrast to that of the armchair system. It is worth mentioning that the absorption spectra of $p_z$ orbital are also greatly enriched by a finite potential, including the enhancement of peaks with lower intensity and the shift of threshold structures, as shown by the dashed red lines in Figs. \ref{Fig3}(Ia) and \ref{Fig3}(IIa). This is mainly attributed to the adjustment of the Fermi level and the deformation of low-lying band edge states under an $E$ field. The first few spectral peaks of the four-orbital and $p_z$-orbital systems are divergent due to the $\sigma$-bands-induced slight difference in Fermi energies. The higher-frequency absorption spectra, which correspond to the vertical transitions of the electronic states away from the Fermi level, become more equivalent for the two models. For example, the ($P_5^{\prime}$ and $\omega_3^{\prime}$) peaks of the armchair GNR (Fig. \ref{Fig3}(Ia)) are located at the same frequency, and so are the further spectral region. Similar behavior is also true for the absorption spectrum of the zigzag GNR, starting from the ($P_3^{\prime}$ and $\omega_3^{\prime}$) peaks (Fig. \ref{Fig3}(IIa)).

\medskip

\begin{figure}[h]
\centering
{\includegraphics[width=0.5\linewidth]{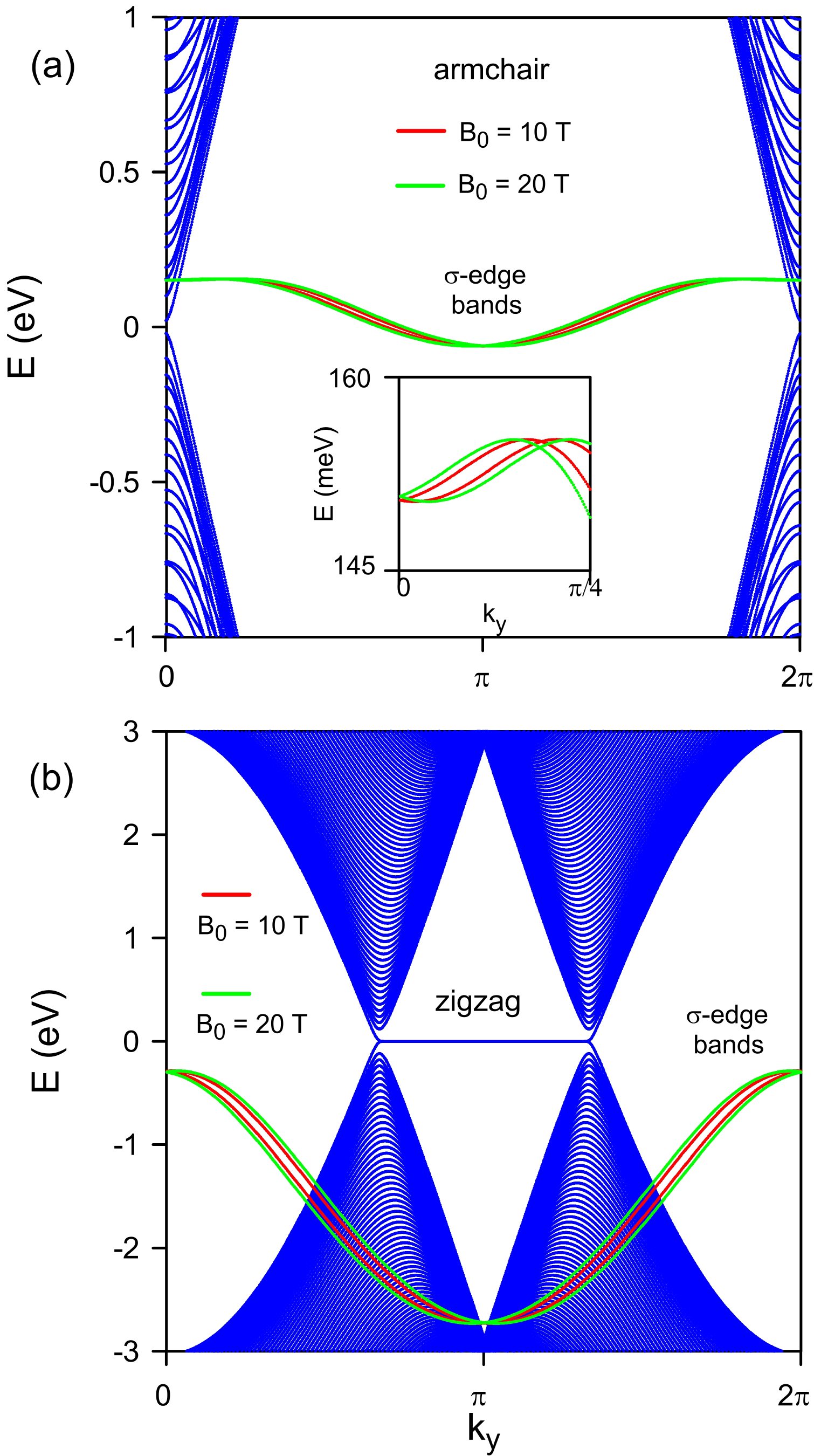}}
\caption{(color online) Calculated $k_y$-dependent Landau levels of GNRs with (a) armchair and (b) zigzag edges for $N = 150$. Here, the doubly degenerate $\sigma$ bands are split by an applied magnetic field, as shown for $B_0 = 10$ T and $20$ T. }
\label{Fig4}
\end{figure}

We now explore in detail the magnetic quantization effect on $\sigma$-edge states by focusing on the (s, $p_x$, $p_y$) orbitals. The typical behavior of the Landau bands (LBs) is sensitively dependent on the ribbon edges. This reflects in the interplay between the magnetic field and the lateral confinement. Figures \ref{Fig4}(a) and \ref{Fig4}(b) show, respectively, the quantized LBs of armchair and zigzag GNRs. An external magnetic field lifts the degeneracy of the $\sigma$-edge bands due to asymmetrical confinement with respect to the edge line for the same direction of two Lorentz forces acting on electrons, leading to two distinct nondegenerate bands. This phenomenon is more perceptible for ribbons with zigzag edges compared with the armchair ones. The $k_y$-dependent LBs of the $\sigma$-edge bands changes greatly with the field strength, referring to the red ($B_0 = 10$ T) and green ($B_0 = 20$ T) curves. The energy splitting between the two $\sigma$-edge bands, $E_g$, strongly depends on $k_y$, the magnetic field strength, and the edge types. The effect of $B_0$ on $E_g$ is stronger for larger magnetic field. Interestingly, the $\sigma$-edge states at $k_y L_y$ = $n\pi$ ($n$ in an integer) remain doubly degenerate without splitting even under the influence of magnetic field. This is because at these special momentum states, the two Lorentz forces with equal magnitude but opposite directions point to the same side of the edge-confinement and therefore they balance each other out. Nevertheless, their energy can still vary with $B_0$. Especially for the armchair-edge system, there exists band crossing behavior in the vicinity of $k_y$ = 0, as illustrated in the zoom-in inset  of Fig. \ref{Fig4}(a). It is noticed that the separation of the two degenerate $\sigma$-edge bands by a magnetic field is unlike that caused by an electric field in terms of energy dispersion and $k_y$-dependent band splitting. Therefore, one might predict a remarkable differences in the influence between the two fields on the optical absorption spectra.
\medskip

\begin{figure}[h]
\centering
{\includegraphics[width=0.5\linewidth]{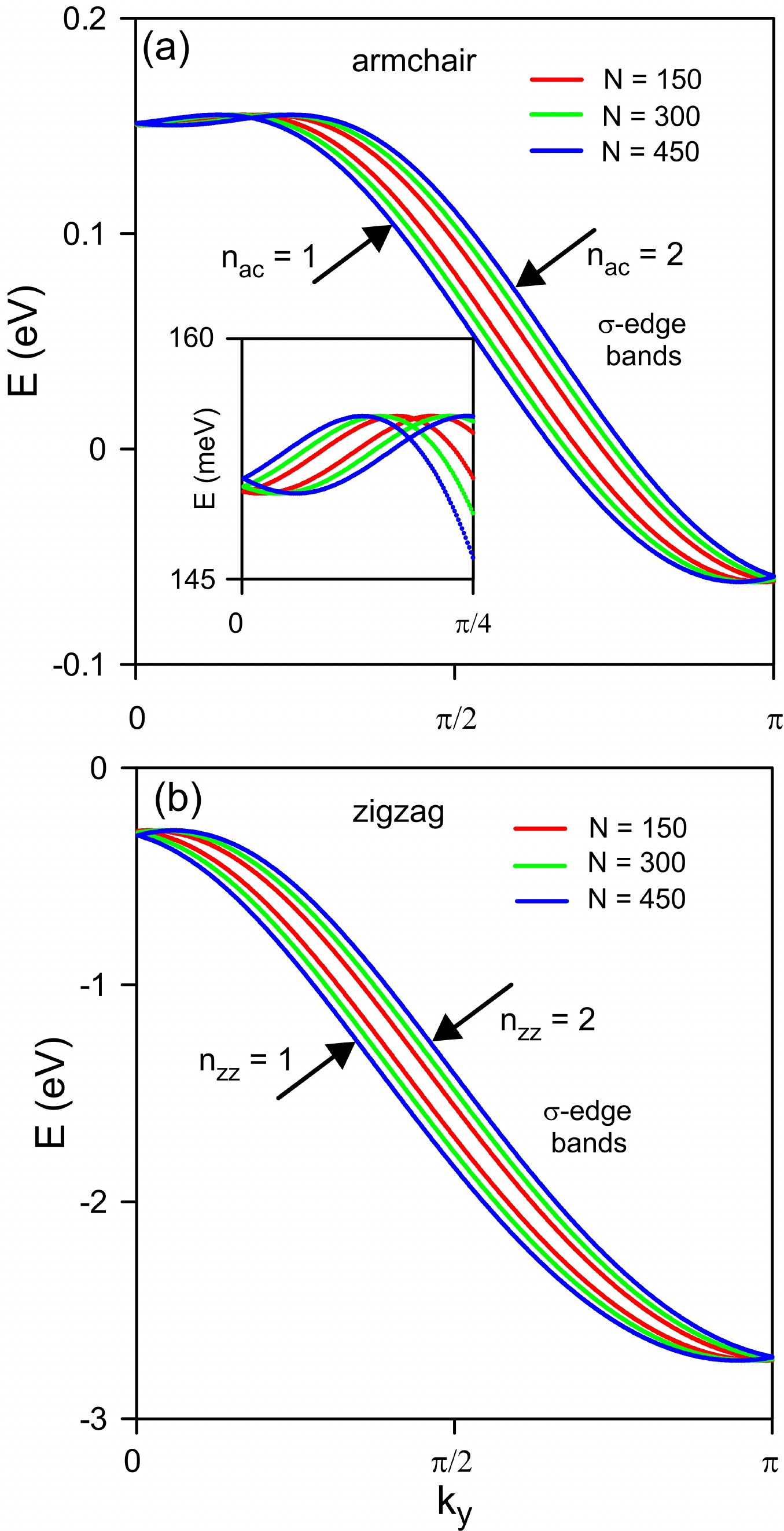}}
\caption{(color online) Calculated dispersion relations of $\sigma$-edge bands as functions of wave number $k_y$ for both (a) armchair and (b) zigzag GNRs. $n_{ac}$/$n_{zz}$ =1 and 2 denote the lower and upper non-degenerate Landau bands of the $\sigma$-edge states. Here, the strong dependence of dispersion on the ribbon width is shown for various values of $N$.}
\label{Fig5}
\end{figure}

The magnetic quantization of the $\sigma$-edge bands is significantly modified by varying the ribbon width. In fact, the inner side of the edge confinement is weakened for increasing ribbon width, leading to a stronger asymmetry or a larger splitting of two $\sigma$ bands. This phenomenon is demonstrated for armchair and zigzag GNRs in Figs. \ref{Fig5}(a) and \ref{Fig5}(b), respectively, for chosen $N$ and $B_0$ = 10 T. The separation between the two degenerate $\sigma$-edge bands, ($n_{ac}$ = 1, $n_{ac}$ = 2) in Fig. 5(a) and ($n_{zz}$ = 1, $n_{zz}$ = 2) in Fig. 5(b), becomes clearer for wider ribbons. Interestingly, the degenerate states at $k_y L_y$ = $n\pi$ are barely affected by the ribbon width, and so is the band crossing of armchair edge systems near $k_y$ = 0 (a close look is inserted in Fig. \ref{Fig5}(a)). The effect of ribbon width on zigzag GNRs in Fig. \ref{Fig5}(b) becomes stronger than that on the armchair system because of an enlarged scale.
It is interesting to notice the $\sigma$-edge bands are moved away from zero energy for both armchair and zigzag GNRs with sufficiently large ribbon width. This conclusion also holds true for the infinite width, i.e., graphene sheet, as demonstrated previously.\,\cite{sigma1, sigma2}
As a matter of fact, the (s, $p_x$, $p_y$) orbitals only have  minor impact on the low-energy band structures of wide GNR systems.
\medskip

\begin{figure}[h]
\centering
{\includegraphics[width=0.7\linewidth]{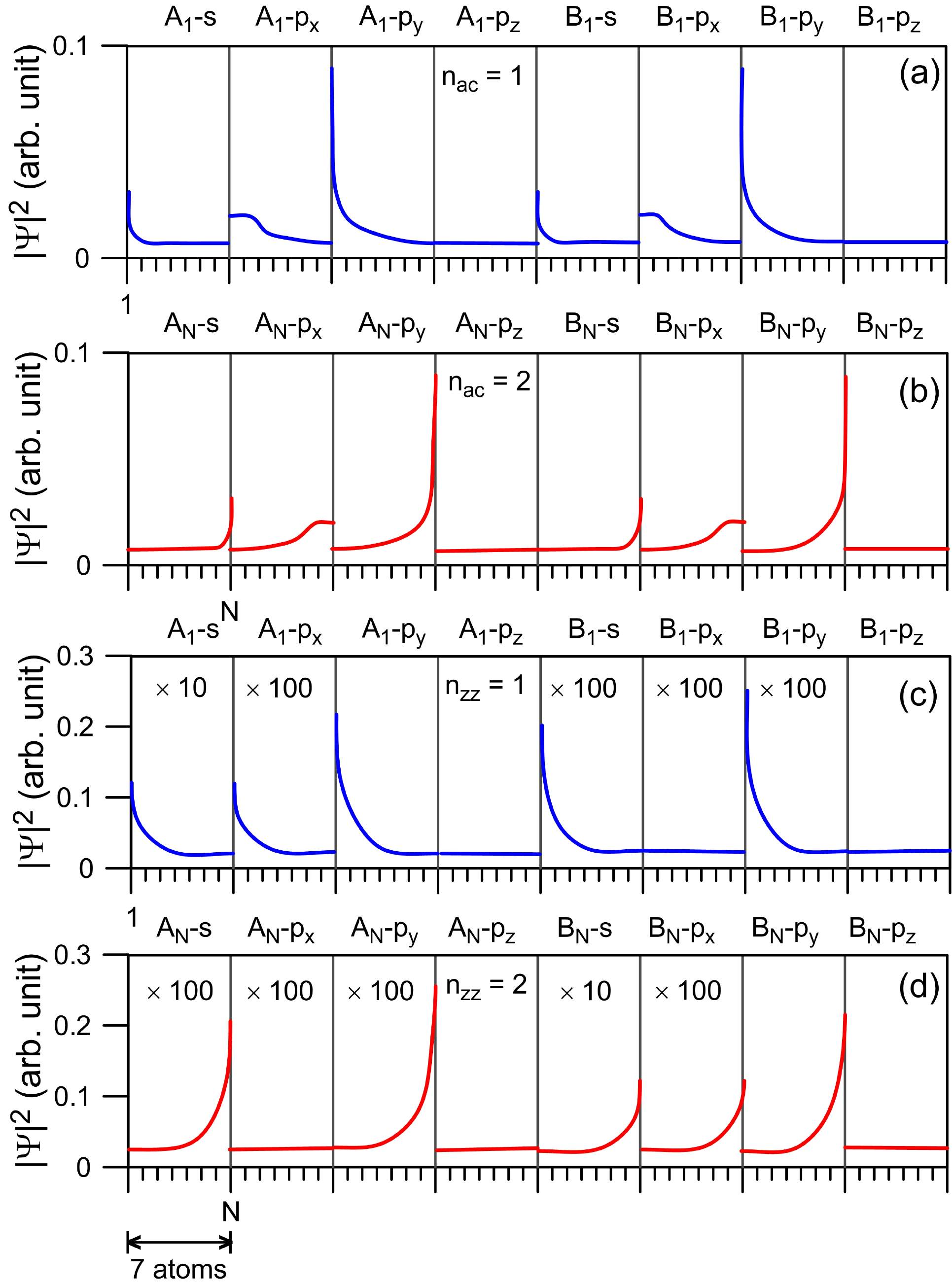}}
\caption{(color online) Calculated position dependence in the probability function $|\Psi|^2$ of $\sigma$-edge bands at $k_y = 2n\pi/L_y$ for (a) $n_{ac}$ = 1 and (b) $n_{ac}$ = 2 LBs of armchair GNRs with $N = 150$. Similar plots for  zigzag GNRs are presented for (c) $n_{zz}$ = 1 and (d) $n_{zz}$ = 2 LBs. Eight displayed columns capture the modes of (s, $p_x$, $p_y$, $p_z$) orbitals
on both A and B sublattices. }
\label{Fig6}
\end{figure}

The dependence of the Landau wave function distribution of the $\sigma$-edge states on the multi-orbitals of A and B sublattices is presented in Figs. \ref{Fig6}(a) and \ref{Fig6}(b) for the armchair GNRs and Figs. \ref{Fig6}(c) and \ref{Fig6}(d) for the zigzag ones. The Landau states of the $\sigma$-edge bands exhibit some unique features which are different from those of the $\pi$ bands. Both the magnetic-field-separated Landau edge bands present the wave function probabilities that are peaked at $k_y L_y$ = 2$n\pi$ bound states. The width of the wave function mode in Fig. \ref{Fig6}(a)  is much smaller than the magnetic length, but it varies with the ribbon width or edge confinement. Each Landau band is attributed to one of the two ribbon edges. The cases with $n_{ac}$ = 1 and $n_{zz}$ = 1 in Figs. \ref{Fig6}(a) and \ref{Fig6}(c) show the finite-amplitude modes localized at the left edge ($k_y$ = 0) while the $n_{ac}$ = 2 and $n_{zz}$ = 2 modes are localized at the right edge ($k_y L_y$ = 2$\pi$). Interestingly, the Landau wave functions of the A atom on the left-hand side of the  ribbon edge  ($n_{ac}$ = 1, $n_{zz}$ = 1) resemble those of the B atom on the right-hand side of the ribbon-edge ($n_{ac}$ = 2, $n_{zz}$ = 2). These unique characteristics are closely related to the asymmetric geometry of GNRs, particularly the positions of the A and B atoms on the two ribbon edges, as illustrated in Figs. \ref{Fig1}(a) and \ref{Fig1}(b).
\medskip

The orbital compositions of the Landau wave functions are not equivalent. The LBs of the $\sigma$ edge bands only have finite modes on the (s, $p_x$, $p_y$) orbitals but vanishing amplitudes on the $p_z$ orbital, which is opposite to those of the $\pi$ bands. This is consistent with the zero field energy dispersion in Fig. \ref{Fig1}.  The role played by each orbital in determining the probability amplitude of the wave function depends on the ribbon edges. As for armchair edge GNRs, the weights of LB wave functions on the (s, $p_x$, $p_y$) orbitals which the A and B atoms are in also contribute of the same order, although the $p_y$ orbitals have slightly higher mode amplitude compared with the other two. On the other hand, the zigzag edge system presents much more visible fluctuation of the LB wave function probability due to the orbitals.  In this case, the wave function of each LB is only dominated by the $p_y$ orbital of either A ($n_{zz}$ = 1) or B ($n_{zz}$ = 2) atom. In contrast, the contribution from $p_x$ orbitals to the Landau wave functions becomes negligible. The dependence of LL wave function distribution on distinct orbitals is critical in understanding the inter-LB optical transition which we discuss in the rest of this paper.
\medskip

\begin{figure}[h]
\centering
{\includegraphics[width=0.7\linewidth]{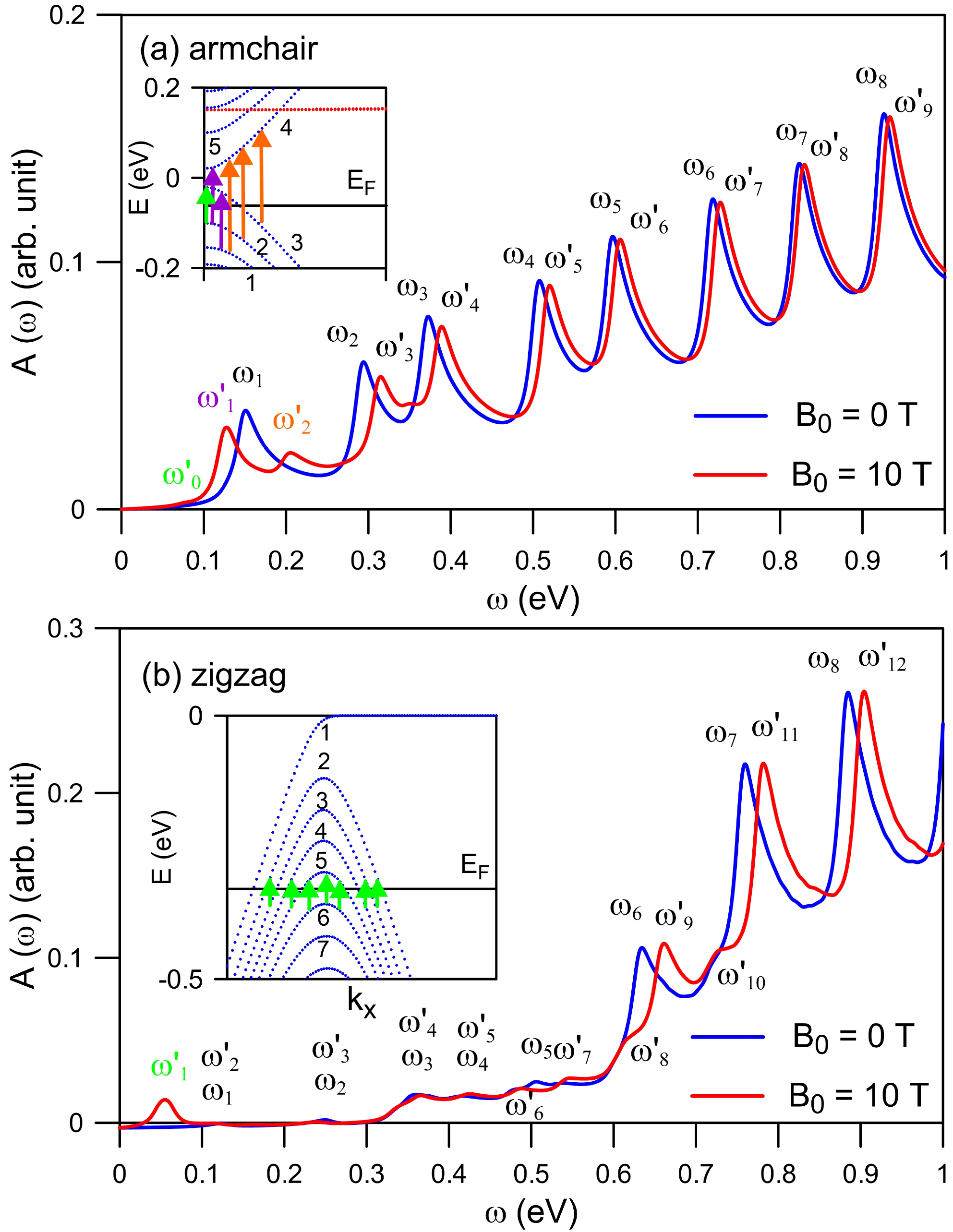}}
\caption{(color online) Calculated optical-absorption spectra with $B_0=0$ and $B_0$ = 10 T are presented for (a) armchair and (b) zigzag GNRs with $N = 150$. The vertical transition channels are displayed in the inserts of (a) and (b).}
\label{Fig7}
\end{figure}

We now focus our attention on the magneto-optical properties of GNRs. For this, we calculate the absorption spectra of GNRs in the absence and presence of a magnetic field, as shown in Figs. \ref{Fig7}(a) and \ref{Fig7}(b) for armchair and zigzag GNRs, respectively. We observe that the low-energy magneto-optical transitions are mainly attributed to the $\pi$ bands. The reason for this is two-fold: (i) the transition between the $\pi$ and $\sigma$ bands is forbidden and (ii) there is no vertical transition between the two degenerate $\sigma$ edge bands. The absorption function exhibits the peak and shoulder-like structures. Overall, the spectral intensity is increased at higher transition frequency $\omega$ due to superposition of high-energy side tails of the density of states for different transitions. A magnetic field of $B_0$ = 10 T only has significant impact on the low frequency peak intensity. As a matter of fact, the higher frequency peak intensity become equivalent to those in the absence of $B_0$. It has been predicted that the spectral intensity could be enhanced by increasing the magnetic field strength or the ribbon width since the absorption peaks are mainly contributed from the $\pi$ bands\,\cite{moptical}.
\medskip

An applied magnetic field can shift the location of peaks, modify the spectral intensity, and create new absorption structures  due to asymmetry in edge channel confinement. The optical thresholds of both the armchair ($\omega^{\prime}_0$ in Fig. \ref{Fig7}(a)) and zigzag GNRs ($\omega^{\prime}_1$ in Fig. \ref{Fig7}(b)) are generated by the finite field. The former is revealed as a relatively low shoulder-like structure which is correlated with the vertical transition between the valence band-edge states of the n = 2 and n = 3 LBs, as indicated by the vertical green arrow in the inset of Fig. \ref{Fig7}(a). On the other hand, the latter appears as a prominent peak, corresponding to the multi-channel vertical transitions among the LBs of n = (1, 2, 3, 4, 5, 6), as illustrated in the inset of Fig. \ref{Fig7}(b). In addition to the thresholds, the other $B_0$-induced extra spectral structures include the special peak $\omega^{\prime}_2$ of armchair GNR and the shoulder-like structures ($\omega^{\prime}_7$, $\omega^{\prime}_8$, $\omega^{\prime}_{10}$) of zigzag GNR. Nevertheless, they are not the dominant structures of the spectra which might not be observable in optical measurements.
\medskip

\begin{figure}[h]
\centering
{\includegraphics[width=0.8\linewidth]{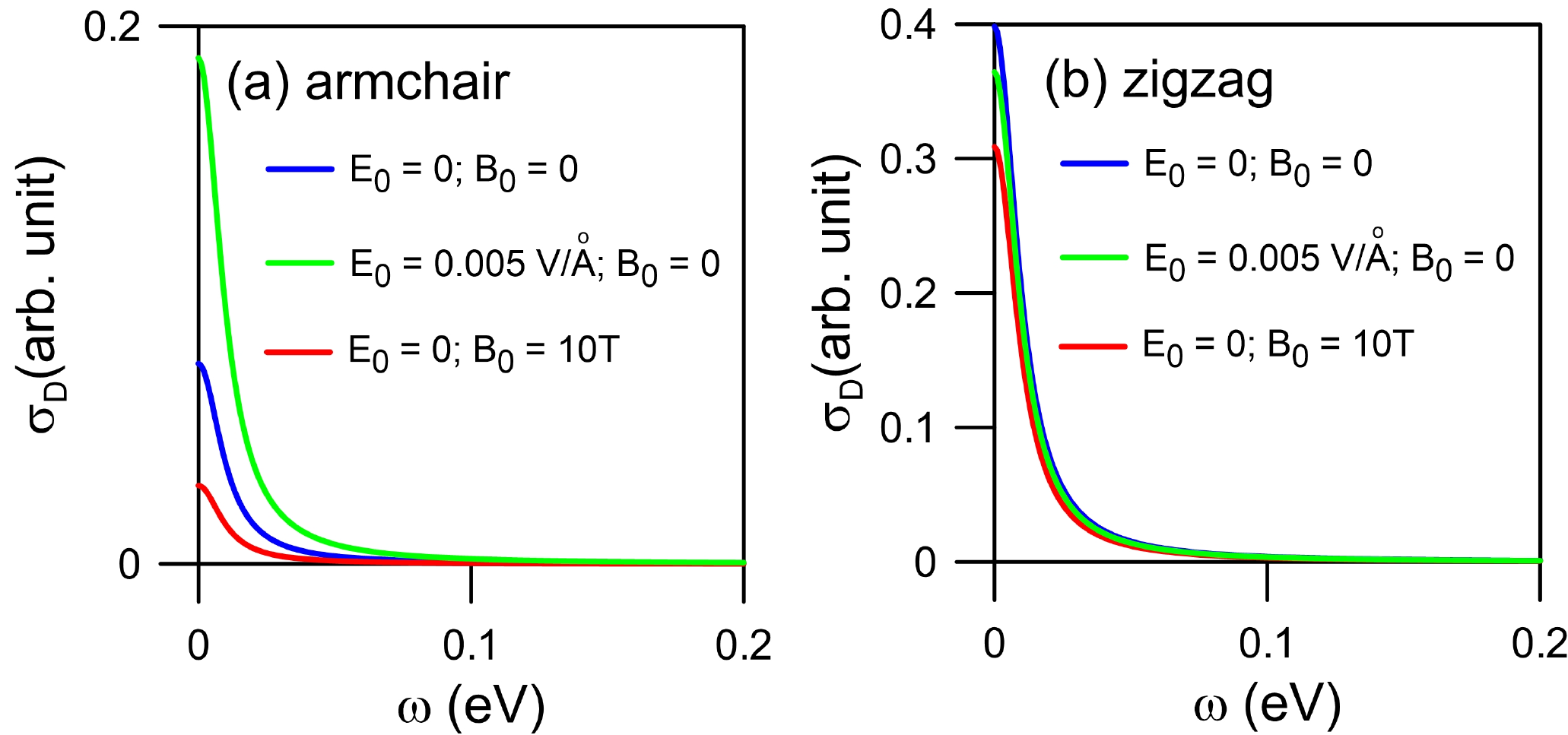}}
\caption{(color online) Calculated Drude contributions to the optical conductivity for (a) armchair and (b) zigzag GNRs for zero and finite electric or magnetic fields.
}
\label{Fig8}
\end{figure}

In addition to the inter-band transition, the Drude conductivity related to the intra-band contribution is an important part of the optical conductivity at low frequencies.  We have employed  Drude's formula for the conductivity based on the Kubo formula from Ref.\ [\onlinecite{Drude1}] to calculate the intra-band conductivity of GNRs.  In order to verify the accuracy of our physics model, we first computed the real and imaginary parts of the conductivity for graphene.  It turns out that our numerical results are in an excellent agreement with the previous experimental and theoretical studies.\cite{Drude2, Drude3}  In Figs.\,8(a) and 8(b), we plot the calculated Drude tails for both armchair and zigzag GNRs at zero electric and magnetic fields,  and at finite electric/magnetic fields.  The intra-band transition only makes remarkable contribution to the optical conductivity at zero frequency.  Both the width and intensity of the Drude tail strongly depend on the number of occupied energy bands across the Fermi surface.  For N = 150 GNRs, the Drude contribution is greatly enhanced by an electric field in the armchair-edge system, which is consistent with the significant shift of $E_F$.
Explicitly, the electric-field-induced shift of Fermi energy leads to the crossing between the Fermi level and a large number of the energy bands, as demonstrated in Fig. 2(a). As a result, the Drude conductivity which comes from the intra-band transitions is much larger for $E_0$ = 0.005 V/\AA \space than that of the zero field.
As for the zigzag GNR, the number of energy bands crossing the Fermi level are comparable for zero-field and $E_0$ = 0.005 V/\AA, therefore, the Drude conductivities of both cases are quite similar, referring to Fig. 8(b).
On the other hand, a finite magnetic field can reduce the Drude conductivity of both the armchair and zigzag GNRs, as shown by the red lines in Figs. 8(a) and 8(b). The field splits the two $\sigma$-edge bands which pulls the Fermi level toward the zero energy where the density of electronic states is lowest. As a result, the intra-band excitation for $B_0$ = 10 T is relatively weaker compared with that of the zero-field case.

\medskip

\section{Concluding Remarks}
\label{sec4}

In this paper, we have investigated the electronic and optical properties of GNRs with armchair and zigzag edges using the TBM as well as the absorption function. By comparing our results for the single $p_z$ and four-orbital Hamiltonian matrices, we were able to understand the significance of the $\sigma$-edge bands for the low energy physics of GNRs. The contribution from  the (s, $p_x$, $p_y$) orbitals to the electronic and optical properties of GNRs is mainly attributed to the emergence of the doubly degenerate $\sigma$-edge bands as well as the shift of the Fermi level.
\medskip

We have also observed that an applied in-plane transverse electric field could alter the energy dispersion which in turn  modifies the inter-band optical transition and Drude conductivity. The presence of an external electric  field results in the splitting of energy bands, distortion of band edge states, and a shift of the Fermi level. These result in crucial changes in the optical conductivity, alteration of the amplitude and frequency of absorption peaks as well as the width and intensity of the Drude tail, and enhancement of shoulder-like structures of the absorption spectra.
\medskip

Finally, we carried out a careful investigation of the quantized Landau bands and the magneto-optical properties, mostly focusing on the effect of the $\sigma$-edge states. The characteristics of the Landau bands depend sensitively on the ribbon edge type, their width, and the magnetic field strength. A magnetic field could split the two degenerate $\sigma$-edge bands to create a $k_x$-dependent band splitting between them. By analyzing the wave functions of these two $\sigma$-edge states, we demonstrated that each of them is attributed to one of the two ribbon sides. The low-energy magneto-absorption spectra, which are governed by the $\pi$ bands, exhibit peak and shoulder-like structures. The spectral intensity is greatly affected by the frequency, ribbon width and edge type, as well as the field strength.
\medskip

\begin{acknowledgements}
D.H. thanks the Air Force Office of Scientific Research (AFOSR)  and the DoD Lab-University Collaborative Initiative (LUCI) program for support. G.G. would like to acknowledge the support from the Air Force Research Laboratory (AFRL through Grant No. 12530960.
\end{acknowledgements}

\end{document}